\documentclass[article,twocolumn,superscriptaddress,showpacs,nofootinbib]{revtex4-1}
\usepackage{amsmath,amssymb,amsfonts,amsthm}
\usepackage{graphicx}
\usepackage{bm}
\usepackage{bbm}
\usepackage{color}
\usepackage[usenames,dvipsnames]{xcolor}
\usepackage{dcolumn}   
\usepackage{epstopdf}
\usepackage[caption=false]{subfig}
\usepackage{setspace}
\usepackage{soul}
\usepackage{color}
\usepackage[colorlinks=true,linkcolor=blue,urlcolor=blue,citecolor=blue]{hyperref}

\begin{document}

 \newcommand{\breite}{1.0} 

\newtheorem{prop}{Proposition}
\newtheorem{cor}{Corollary}

\newcommand{\be}{\begin{equation}}
\newcommand{\ee}{\end{equation}}

\newcommand{\bea}{\begin{eqnarray}}
\newcommand{\eea}{\end{eqnarray}}
\newcommand{\lt}{<}
\newcommand{\gt}{>}

\newcommand{\Reals}{\mathbb{R}}     
\newcommand{\Com}{\mathbb{C}}       
\newcommand{\Nat}{\mathbb{N}}       

\newcommand{\mkch}[1]{{\color{BrickRed} #1}}

\newcommand{\id}{\mathbboldsymbol{1}}

\newcommand{\Real}{\mathop{\mathrm{Re}}}
\newcommand{\Imag}{\mathop{\mathrm{Im}}}

\def\O{\mbox{$\mathcal{O}$}}   
\def\F{\mathcal{F}}			
\def\sgn{\text{sgn}}

\newcommand{\deo}{\ensuremath{\Delta_0}}
\newcommand{\dea}{\ensuremath{\Delta}}
\newcommand{\ak}{\ensuremath{a_k}}
\newcommand{\ad}{\ensuremath{a^{\dagger}_{-k}}}
\newcommand{\sx}{\ensuremath{\sigma_x}}
\newcommand{\sz}{\ensuremath{\sigma_z}}
\newcommand{\spl}{\ensuremath{\sigma_{+}}}
\newcommand{\smi}{\ensuremath{\sigma_{-}}}
\newcommand{\alk}{\ensuremath{\alpha_{k}}}
\newcommand{\bk}{\ensuremath{\beta_{k}}}
\newcommand{\ok}{\ensuremath{\omega_{k}}}
\newcommand{\vd}{\ensuremath{V^{\dagger}_1}}
\newcommand{\vi}{\ensuremath{V_1}}
\newcommand{\vo}{\ensuremath{V_o}}
\newcommand{\zc}{\ensuremath{\frac{E_z}{E}}}
\newcommand{\xc}{\ensuremath{\frac{\Delta}{E}}}
\newcommand{\xd}{\ensuremath{X^{\dagger}}}
\newcommand{\aok}{\ensuremath{\frac{\alk}{\ok}}}
\newcommand{\tpw}{\ensuremath{e^{i \ok s }}}
\newcommand{\tpe}{\ensuremath{e^{2iE s }}}
\newcommand{\tmw}{\ensuremath{e^{-i \ok s }}}
\newcommand{\tme}{\ensuremath{e^{-2iE s }}}
\newcommand{\epls}{\ensuremath{e^{F(s)}}}
\newcommand{\emis}{\ensuremath{e^{-F(s)}}}
\newcommand{\epl}{\ensuremath{e^{F(0)}}}
\newcommand{\emi}{\ensuremath{e^{F(0)}}}

\newcommand{\lr}[1]{\left( #1 \right)}
\newcommand{\lrs}[1]{\left( #1 \right)^2}
\newcommand{\lrb}[1]{\left< #1\right>}
\newcommand{\nbt}{\ensuremath{\lr{ \lr{n_k + 1} \tmw + n_k \tpw  }}}

\def\beq{\begin{equation}}
\def\eeq{\end{equation}}
\def\bea{\begin{eqnarray}}
\def\eea{\end{eqnarray}}

\newcommand{\om}{\ensuremath{\omega}}
\newcommand{\dw}{\ensuremath{\Delta_0}}
\newcommand{\wbp}{\ensuremath{\omega_0}}
\newcommand{\dv}{\ensuremath{\Delta_0}}
\newcommand{\vbp}{\ensuremath{\nu_0}}
\newcommand{\vplus}{\ensuremath{\nu_{+}}}
\newcommand{\vminus}{\ensuremath{\nu_{-}}}
\newcommand{\wplus}{\ensuremath{\omega_{+}}}
\newcommand{\wminus}{\ensuremath{\omega_{-}}}
\newcommand{\uv}[1]{\ensuremath{\mathbf{\hat{#1}}}} 
\newcommand{\abs}[1]{\left| #1 \right|} 
\newcommand{\avg}[1]{\left< #1 \right>} 
\let\underdot=\d 
\renewcommand{\d}[2]{\frac{d #1}{d #2}} 
\newcommand{\dd}[2]{\frac{d^2 #1}{d #2^2}} 
\newcommand{\pd}[2]{\frac{\partial #1}{\partial #2}}
\newcommand{\pdd}[2]{\frac{\partial^2 #1}{\partial #2^2}}
\newcommand{\pdc}[3]{\left( \frac{\partial #1}{\partial #2}
 \right)_{#3}} 
\newcommand{\ket}[1]{\left| #1 \right>} 
\newcommand{\bra}[1]{\left< #1 \right|} 
\newcommand{\braket}[2]{\left< #1 \vphantom{#2} \right|
 \left. #2 \vphantom{#1} \right>} 
\newcommand{\matrixel}[3]{\left< #1 \vphantom{#2#3} \right|
 #2 \left| #3 \vphantom{#1#2} \right>} 
\newcommand{\grad}[1]{{\nabla} {#1}} 
\let\divsymb=\div 
\renewcommand{\div}[1]{{\nabla} \cdot \boldsymbol{#1}} 
\newcommand{\curl}[1]{{\nabla} \times \boldsymbol{#1}} 
\newcommand{\laplace}[1]{\nabla^2 \boldsymbol{#1}}
\newcommand{\vs}[1]{\boldsymbol{#1}}
\let\baraccent=\= 

\newcommand{\sg}[1]{{\color{red} #1}}
\newcommand{\dah}[1]{{\color{blue} #1}}
\newcommand{\mk}[1]{{\color{cyan} #1}}
\newcommand{\ka}[1]{{\color{magenta} #1}}
\newcommand{\ea}[1]{{\color{olive} #1}}


\title{Rare-region effects and dynamics near the many-body localization transition}

\author{Kartiek Agarwal}
\affiliation{Department of Electrical Engineering, Princeton University, Princeton NJ 08544, USA}%
\author{Ehud Altman}
\affiliation{Department of Physics, University of California, Berkeley, CA 94720, USA}
\author{Eugene Demler}
\affiliation{Department of Physics, Harvard University, Cambridge MA 02138, USA}
\author{Sarang Gopalakrishnan}
\affiliation{Department of Engineering Science and Physics, CUNY College of Staten Island, Staten Island, NY 10314 USA}
\author{David A. Huse}
\affiliation{Physics Department, Princeton University, Princeton, NJ 08544, USA}
\author{Michael Knap}
\affiliation{Department of Physics, Walter Schottky Institute, and Institute for Advanced Study, Technical University of Munich, 85748 Garching, Germany}

\date{\today}
\begin{abstract}
The low-frequency response of systems near the many-body localization phase transition, on either side of the transition, is dominated by contributions from rare regions that are locally ``in the other phase'', i.e., rare localized regions in a system that is typically thermal, or rare thermal regions in a system that is typically localized. Rare localized regions affect the properties of the thermal phase, especially in one dimension, by acting as bottlenecks for transport and the growth of entanglement, whereas rare thermal regions in the localized phase act as local ``baths'' and dominate the low-frequency response of the MBL phase. We review recent progress in understanding these rare-region effects, and discuss some of the open questions associated with them: in particular, whether and in what circumstances a single rare thermal region can destabilize the many-body localized phase. 
\end{abstract}
\maketitle

\section{Introduction}

A basic assumption of statistical mechanics is that isolated, but internally interacting many-body physical systems, ``thermalize'': i.e., starting from generic initial conditions, they tend towards a state in which entropy is maximized, subject to constraints due to a small number of conservation laws. Thus, equilibrium states can be statistically described in terms of a few parameters, i.e., the conserved densities or their associated Lagrange multipliers~\cite{rigolGGE}, such as the temperature or the chemical potential. This ``thermalization'' is expected to occur generically in classical systems with chaotic dynamics.  In quantum-mechanical systems, the expectation that such isolated systems will thermalize is captured by the ``eigenstate thermalization hypothesis'' (ETH)\cite{deutsch_quantum_1991, srednicki_chaos_1994, rigol_thermalization_2008}.  The ETH states that the eigenstates of a generic many-body quantum system are locally ``thermal,'' in the sense that the reduced density matrix of a sufficiently small {\it subsystem} is the same when the full system is in an eigenstate as when the full system is in any other thermal equilibrium state, such as the Boltzmann distribution.  Numerical evidence supports the ETH in a variety of quantum systems~\cite{rigolGGE, rigol_thermalization_2008, kih_outliers}, although it is known to fail in certain experimentally relevant special cases, e.g., a one-dimensional Bose (or Fermi) gas with short-range interactions~\cite{newtonscradle, rigolGGE, rigol_thermalization_2008}.  Such special cases, corresponding to integrable (and thus non-chaotic) dynamics, were thought to be fine-tuned \emph{points} rather than stable dynamical phases.  However, in recent years, a stable, nonthermalizing specifically quantum \emph{phase} known as the `many-body-localized' (MBL) phase, has been predicted to exist in certain systems~\cite{BAA, Gornyi, OH, NH, AV}.  The MBL phase is not known to have any direct classical equivalent~\cite{oph, Basko}, and unlike traditional integrable systems is, moreover, robust against generic local perturbations to the system's Hamiltonian, so is not fine-tuned. 

In addition to its conceptual importance, the question of how and whether or not an isolated quantum system thermalizes has been a focus of current experiments realizing ``synthetic quantum matter''. That such questions can be probed experimentally at all is the result of advances in preparing, controlling and measuring such systems in a variety of platforms, including ultracold atomic~\cite{bloch_many-body_2008,langenultracoldnoneq}, trapped ion systems~\cite{blatt_quantum_2012,trappedionmonroe}, superconducting qubit arrays~\cite{houck2012chip}, NV-centers~\cite{kucsko2016critical} etc. These developments bring the investigation of out-of-equilibrium many-body quantum dynamics within experimental reach; and, indeed, both the failure of thermalization in integrable one-dimensional quantum systems~\cite{paredes_tonks-girardeau_2004, newtonscradle, gring_relaxation_2012} and the presence of MBL regimes have been experimentally demonstrated~\cite{Schreiber15, kondov_Disorder_2015, smith2015, Bordia16, Hild16, bordia_periodically_2016}.

In recent years, a detailed phenomenological understanding has emerged for the fully MBL phase, in which all eigenstates are localized. This phase supports an extensive set of \emph{localized} integrals of motion~\cite{Vosk2013,spa1, husephenom, spa2, ros2015integrals, jzi_long} (termed LIOMs or ``l-bits''), and certain quantum correlations can retain memory of their initial state even at infinitely late times~\cite{palmbl}. The MBL phase resembles noninteracting Anderson insulators in some ways (e.g., spatial correlations decay exponentially, and eigenstates have area-law entanglement~\cite{bauerarealaw}). However, there are also important distinctions in entanglement dynamics~\cite{znidaric_many-body_2008,bardarsonentanglementgrowth}, dephasing~\cite{serbyninterferometric,spa_quench,vasseur_revivalsMBL}, linear~\cite{mbmott} and nonlinear~\cite{ldm, adh2014, ppha, kns, kozarzewski2016, rehn2016, gkd} response, and the entanglement spectrum~\cite{geraedtsspectrum, smap}. These developments (reviewed in Refs.~\cite{NH,AV,imbrie2016review}) are not the focus of the present review.

The present review is concerned with rare-region effects associated with the many-body localization phase transition. The presence of thermodynamic and dynamic singularities due to rare regions in systems with quenched randomness and a phase transition was first pointed out by Griffiths~\cite{griff} and by McCoy~\cite{mccoy}.  We will follow current usage and call these effects ``Griffiths effects''. They are one type of precursor to the phase transition: rare, large but finite ``inclusions'' of the other phase, i.e., regions in which (because of highly atypical configurations of the quenched randomness) the parameters of the system are locally such that it appears to be in the other phase.  Such Griffiths effects have been studied extensively at conventional thermal~\cite{griff,mccoy} and quantum phase transitions~\cite{damlermotrunichhuse1,damlermotrunichhuse2}; see e.g. Ref.~\cite{vojta_review} for a review. At some phase transitions, such as the quantum phase transition from a magnetically disordered to a magnetically ordered phase in the random transverse-field Ising spin chain, the leading divergences near the phase transition are due to such Griffiths effects~\cite{dsf}. 

Here, we shall be concerned with Griffiths effects near the \emph{unconventional} dynamic quantum phase transition between the thermal and MBL phases. The MBL phase transition is still poorly understood, so whether these Griffiths effects are the dominant precursors to the phase transition, on either side of it, is still not clear. 
In the thermal phase of a system with quenched randomness there may be rare regions that are more disordered and thus locally MBL.  Ultimately, these rare regions do thermalize due to being surrounded by thermal regions, which act as a ``bath,'' but their slow dynamics can dominate the long-time or low-frequency dynamics of such a system~\cite{VHA,Agarwal,gadhk}. These effects are most dramatic in one dimension, where such inclusions of the MBL phase can bottleneck transport, making the diffusion constant vanish even within the thermal phase, as has been seen in some numerical studies~\cite{BarLev,znidaric_diffusive_2016}.
 On the other hand, in the MBL phase of a system with quenched randomness there may be rare regions that are less disordered and thus locally thermalizing~\cite{mbmott}. 
Ultimately, at least in one-dimensional systems with short-range interactions, these rare regions get ``localized'' because they are effectively finite~\cite{mbmott,VHA,imbrieproof,drh}: thus the spectral features of an inclusion become discrete on frequency scales small compared with the level spacing of the inclusion. 
In addition to rare regions of the disorder potential, the MBL phase has rare regions of the \emph{state}~\cite{drhms}, in which the conserved densities have atypical values (such ``rare regions'' would of course not be dynamically stable in the thermal phase). If the system has a putative many-body mobility edge separating MBL and thermal energy ranges, a state that is globally deep in the MBL energy range can have a local energy fluctuation that takes it to, or beyond, the energy density corresponding to the many-body mobility edge. Unlike rare regions of the disorder potential, rare regions of the state do not depend on the existence of quenched randomness, and could also apply to nonrandom quasiperiodic systems that show MBL~\cite{iyerquasiperiodicmbl}.

Within each phase, large inclusions of the other phase are sparse, so in the first approximation they can be treated as noninteracting. Therefore it may suffice to (i) count inclusions, and (ii) understand the influence of a \emph{single, isolated} wrong-phase inclusion as a function of its size and properties.
For MBL inclusions in the thermal phase, both steps are relatively straightforward: assuming spatially uncorrelated disorder and a one-dimensional geometry, the density of localized inclusions of linear dimension $L$ is exponentially small in $L$.  Further, the transit time $t$ for information or particles to traverse a localized inclusion of length $L$ can be seen to be $t(L, \zeta) \sim e^{L/\zeta}$, with a finite decay length $\zeta$~\cite{VHA,Agarwal}.  Thus, inclusions with transit times $\sim t$ have length $L \propto \zeta\log t$, and the density of such inclusions falls off as a power of $t$ with an exponent that varies throughout the thermal phase; as we discuss in Sec.~\ref{sec:GriffithsinThermal}, these continuously varying power laws can dominate various physical properties in the thermal phase near the MBL transition.

The case of a thermal inclusion in the MBL phase is more subtle, and currently not fully understood. There is an essential dichotomy between the thermal and MBL phases: the thermal phase is incoherent, while the MBL phase is coherent, and in the ``battle'' between coherence and decoherence, decoherence always has the upper hand.  Thus thermal inclusions within the MBL phase are potentially much more potent and their effects more far-reaching as compared to MBL inclusions within the thermal phase.  
This is because a thermal inclusion can ``grow'' by acting as a bath for degrees of freedom near it, and the processes that govern and limit this growth are not entirely clear at present.
Indeed, there are scenarios in which a single rare thermal region, of either the disorder potential or the state, causes a nonperturbative instability of the many-body localized phase~\cite{drh,drhms}. In particular, such scenarios have been constructed to argue against the existence of many-body mobility edges or of a MBL phase in high-dimensional systems.  However, even if these scenarios are correct, ``MBL glass'' regimes that are effectively many-body localized on all accessible time and length scales will still exist, they just may be separated from the thermalizing regimes by a crossover rather than by a sharp phase transition.

 In what follows, we shall first review what is understood about Griffiths effects within the thermal phase (Sec. \ref{sec:GriffithsinThermal}) and then discuss rare-region effects on the MBL side of the transition (Sec. \ref{sec:GriffithsinMBL}); and finally turn to the renormalization-group theories of the transition (Sec. \ref{sec:MBLtrans}).
We shall focus more on one-dimensional systems, although we shall also discuss the implications for higher-dimensional MBL systems, including the question of to what extent such systems even exist.

\section{Thermal Phase}
\label{sec:GriffithsinThermal}
\begin{figure}
\includegraphics[width=3.3in]{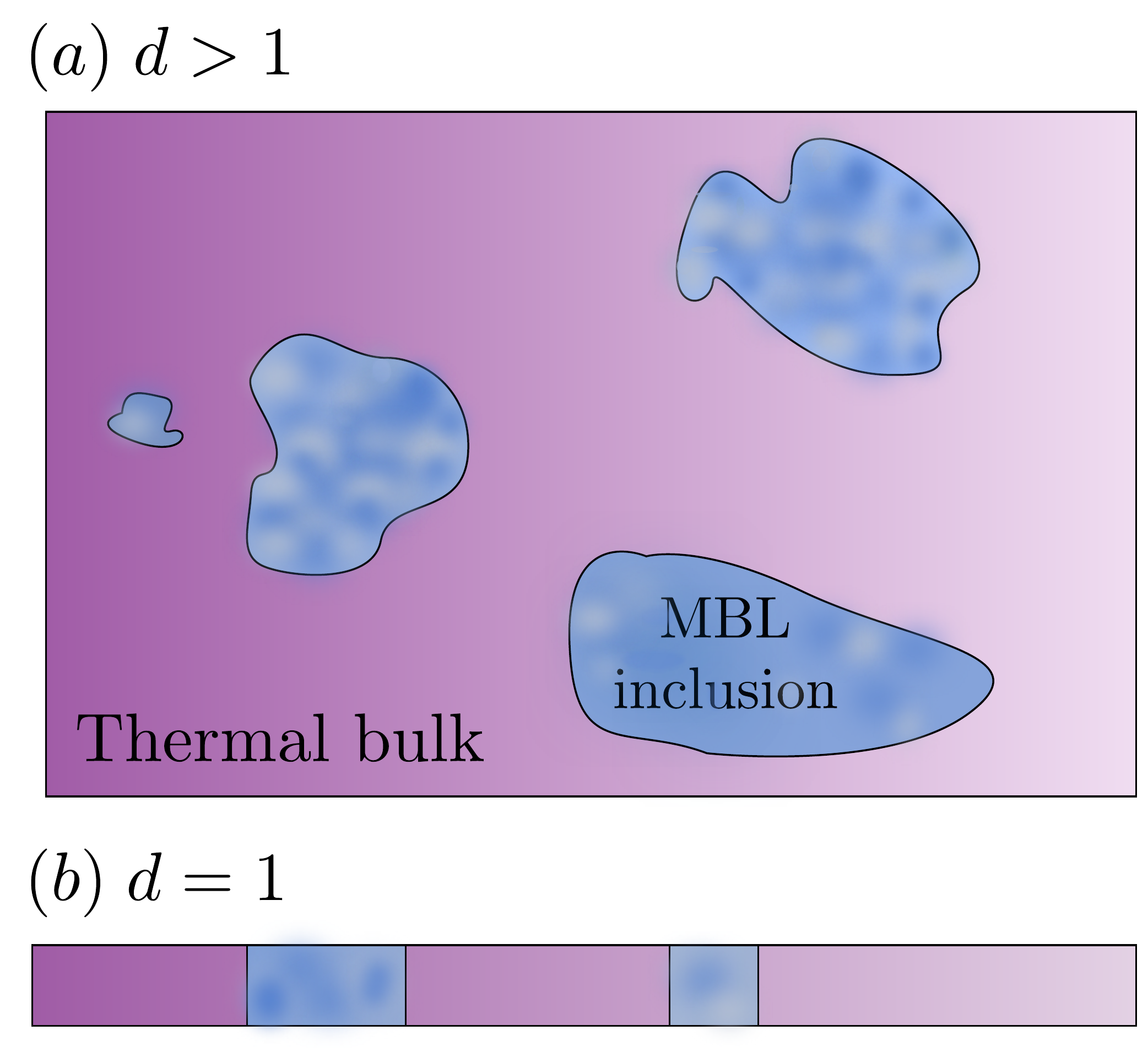}
\caption{Griffiths effects in the thermal phase and dimensionality. (a) In dimensions $d > 1$, transport, e.g., a heat current induced by a small temperature gradient (shading of the background), can go around slow insulating regions (blue speckled regions). (b) For $d = 1$, rare insulating regions can act as ``bottlenecks'', since any transport current has to pass through them.}
\label{fig:bottleneck}
\end{figure}

On the thermal side of the MBL transition, rare large locally insulating regions can significantly alter the transport properties of the system\footnote{Here, a distinction must be made between systems with quenched randomness and systems with deterministic quasi-periodic potentials or fields.  The latter do not harbor Griffiths regions due to rare fluctuations of the disorder, though rare random fluctuations of the state may give rise to similar effects.} (Fig.~\ref{fig:bottleneck}). This is especially true in one dimension, where insulating regions can act as bottlenecks that limit transport across the system. Thus, in one dimension even the \emph{typical} dynamics\footnote{Griffiths effects can lead to broad distributions (over many disordered samples) of various dynamical observables. Thus, it is important to distinguish between \emph{average} and \emph{typical} properties.  Experimentally, the average result is obtained after measurements over an exhaustive range of samples that includes samples containing the rarest of rare insulating inclusions, and the average may be dominated by rare samples; the typical result is the more likely outcome of the measurement in any one particular sample.  Mathematically, the typical value may be defined as the median of the distribution.} of the system can be dominated by rare-region effects. In higher dimensions, transport and operator spreading can bypass rare insulating regions, so global transport and \emph{typical} relaxation are not dominated by rare-region effects.  However, certain spatially-averaged properties can still in principle be dominated by rare-region effects: local observables deep \emph{inside} a rare insulating region can take a very long time to relax, so the slowest relaxation in the system can be dominated by rare-region effects.

\subsection{Response of a single inclusion}\label{STI}

Suppose the MBL transition is governed by a global control parameter $\delta$ (e.g., disorder or interaction strength), such that when the typical value of $\delta < 0$ the system is in the thermal phase, and otherwise the system is MBL.
For a sufficiently large but finite region of a system, one can meaningfully define a \emph{local} value of $\delta$, denoted $\delta_i$, which determines whether the inclusion is \emph{locally} in the thermal or MBL phase.
We now consider a $d$-dimensional thermal system with an approximately spherical MBL inclusion of radius $R_0$ at the origin: i.e., the system is locally MBL (with local control parameter ${\delta}_i\geq 0$) for $R < R_0$, and thermal (with the typical value $\delta<0$ for its control parameter) for $R > R_0$. Following Ref.~\cite{VHA}, we can split up the Hamiltonian into $\hat{H}_{R < R_0} + \hat{H}_{R \agt R_0} + \hat{H}'$, where $\hat{H}'$ consists of boundary terms that connect the subregions. We diagonalize $\hat{H}_{R < R_0}$ by local integrals of motion or ``l-bits''\cite{husephenom}, in terms of which it has the form $\hat{H}_{R < R_0} = h_i \hat{\tau}^z_i + J_{ij} \hat{\tau}^z_i \hat{\tau}^z_j + \ldots$. The boundary terms in $\hat{H}'$ can be expressed as $\hat{A}_{R < R_0} \hat{B}_{R \geq R_0}$, where $\hat{A}$ and $\hat{B}$ are local operators living respectively inside and outside the inclusion. The operator $\hat{A}$ can be expanded in terms of l-bits: the coefficients will decay exponentially with distance from the boundary. 
Meanwhile, 
we can model the thermal region as an infinite heat bath.
Each l-bit inside the inclusion is coupled, by terms in $H'$, to this thermal bath, with a coupling constant that falls off exponentially with its distance from the boundary. The longest relaxation timescale is then 
\beq\label{tR}
t(R_0) \propto \exp(R_0/\zeta)
\eeq
where $\zeta (\delta_i)$ is the characteristic decay length in the inclusion of the relaxational interactions with the thermal bath. This conclusion is precisely what one might have expected on intuitive grounds for a generic insulator.

The definition of Griffiths regions in a system close to the MBL critical point requires more care. In this regime, even the thermal bulk only appears thermal when probed on lengths scales larger than the correlation length $\xi$, expected to diverge at the critical point, and on timescales that are long compared with its correlation time $\tau(\xi)$ [which is thought to go as $\exp(\xi/\zeta)$ ~\cite{VHA, PVP}]. Hence the boundary between the Griffiths region and the thermal bulk is smeared over the scale $\xi$ and for isolated Griffiths regions to be well defined they must be of size $R_0\gg\xi$.
The critical behavior, seen on time scales shorter than $\tau(\xi)$ is described in the RG approach in terms of {\em strongly coupled} insulating and thermal regions present on all scales below $\xi$. For a sufficiently large inclusion we expect the asymptotic size-dependence of the decay rate to remain exponential, $t(R_0) \propto \exp(R_0/\zeta)$, but the sub-exponential prefactor will presumably be renormalized away from its Golden-Rule value as the thermal bulk approaches the MBL transition.

\subsection{Counting inclusions: the role of dimensionality}

Equipped with the previous result, we can estimate the density of insulating inclusions in the thermal phase. An inclusion is considered ``insulating'' on timescale $t$ if information takes longer than $t$ to escape from the inclusion. Smaller inclusions are fully relaxed to their local environment on timescale $t$ and are considered ``conducting'' on that timescale. From the previous discussion, an inclusion that is insulating on timescale $t$ must be of radius $R_0 \agt \zeta(\delta_i) \log t$, where $\delta_i$ is the value of the control parameter inside the inclusion. The probability of such an inclusion can be argued, on the grounds of large-deviations theory, to be $p \sim e^{- r (\delta, \delta_i) L^d}$, where $r \geq 0$ is a rate function that depends on both $\delta$, the value of the control parameter in the bulk phase, and $\delta_i$, the respective value in the inclusion; $r(\delta,\delta_i)=0$ only when $\delta_i=\delta$.

In one dimension, near the MBL critical point, the occurrence of inclusions falls off exponentially with their size $R_0$ as $p(R_0)\sim e^{-R_0/\xi}$. Concomitantly, the relaxation time inside (or across) these inclusions is exponentially large in $R_0$, $t\sim e^{R_0/\zeta}$. As a result, the (spatial) density of inclusions that are insulating on time $t$ is a power-law $t^{-1/z}$. Correspondingly the system develops a power-law distribution of relaxation time scales $p(t)\sim t^{-(z+1)/z}$. The above arguments suggest that near the MBL phase transition $z$ varies continuously as $z=\xi/\zeta$ and hence diverges together with $\xi$ at the critical point. This conclusion is supported by the renormalization group treatments \cite{VHA,PVP}, which also suggest, as does numerical work \cite{palmbl}, that transport through a {\it critical} region of size $R_0$ occurs on a timescale $t(R_0) \sim \exp(R_0/\zeta_c)$.

In higher dimensions, the density of inclusions that are insulating at time $t$ falls off as $\exp(- \alpha \log^d t)$, which is faster than a power law. Thus, in higher dimensions, Griffiths effects in response are generically sub-leading in the thermal phase to the power-law hydrodynamic long-time tails that dominate the late-time response of many observables. The only exceptions to this are systems with no conserved quantities at all (e.g., certain driven systems), for which hydrodynamic long-time tails are absent, or very specific observables that do not have long-time tails~\cite{gadhk}.

\subsection{Griffiths effects in one dimension: density and current response}

In one dimension, Griffiths effects lead to power-law singularities in the time- or frequency-dependence of various dynamical observables that may dominate over hydrodynamic power laws, provided $z$ is large enough.  We now briefly review what is believed about the behavior of various observables in the Griffiths regime of the thermal phase.  Our considerations here are generic and apply to any interacting system with quenched randomness and at any nonzero temperature, provided that one goes to sufficiently late times. Recall that rare regions contribute to dynamics in two distinct ways in one dimension: first, the dynamics inside a rare region is slow, and these slowly relaxing regions can dominate the late-time behavior of spatially averaged quantities; and second, rare insulating regions can act as bottlenecks for transport, operator spreading and the growth of entanglement. The former of these effects is closely analogous to conventional Griffiths effects, e.g., in disordered magnets, whereas the latter is somewhat different.

\emph{Elastic vs. inelastic processes}.---There are in principle two ways for particles or information to go through an MBL inclusion: they can be transmitted ``elastically'' (by tunneling through the inclusion while making only virtual changes to the state of the l-bits in the inclusion) or ``inelastically'' (by processes that produce real changes in the state of the inclusion). Elastic processes give transport across an inclusion but do not thermalize the inclusion itself, whereas inelastic processes thermalize the inclusion.  At the Golden Rule level of analysis, it seems that inelastic processes dominate~\cite{gadhk} for a large enough inclusion (since the matrix element for going fully across the inclusion is exponentially suppressed relative to the matrix element for going halfway into the inclusion).  If instead it is the case that elastic processes can be faster than inelastic for long inclusions, then we will have two dynamic exponents $z$ with the larger one governing the slower relaxation of the l-bits within the inclusion.

\emph{Decay of generic spatially averaged autocorrelation functions}.---Generic local autocorrelation functions in the MBL phase remain nonzero in the infinite-time limit. Thus, autocorrelation functions of operators \emph{inside} a MBL inclusion of size $L$ will remain of order unity at times shorter than the decay timescale $t(L)$ [Eq.~\eqref{tR}]. The contribution of these inclusions to spatially averaged autocorrelation functions is set by their density, and goes as $t^{-1/z}$. 
Thus we expect generic autocorrelation functions to decay as $t^{-1/z}$ or slower. Certain autocorrelation functions, such as that of the current, that are forced by symmetry (in the case of current, time-reversal symmetry) to vanish at long times in the MBL phase are exceptions to such a decay law. 

\emph{Autocorrelation functions of conserved densities; subdiffusion}.---As discussed above, in one dimension inclusions can also affect the \emph{typical} behavior of autocorrelation functions by acting as transport bottlenecks. We now discuss this effect for the case of density relaxation. Let us first consider density relaxation between two good thermal regions of size $L$ separated by a bottleneck, across which particles randomly hop with a rate $\Gamma$. Then the timescale for density to relax across the bottleneck is $\sim L/\Gamma$, i.e., it is the time required for essentially all the particles in the thermal regions to ``forget'' their initial condition. Let us now consider a region of size $L$ and estimate the time it takes density to relax across it. The slowest expected bottlenecks in such a region are those with density $t^{-1/z}\sim1/L$, i.e we expect to find about one such bottleneck in the region; these bottlenecks therefore have timescale $t\sim L^{z}$. Consequently, the time it takes density to relax across $L$ is $t_\text{tr}(L) \sim L^{1 + z}$. Inverting this relation, we arrive at the relation
\beq
L \sim t_\text{tr}(L)^\beta, \quad \beta = 1/(1 + z)
\eeq
Note that these bottlenecks only dominate transport when $z > 1$, i.e., when $\beta < 1/2$.  In this regime, the system exhibits subdiffusive dynamics: its d.c. diffusion constant and therefore its d.c. conductivity vanish, although the system still thermalizes. 

\begin{center}
\begin{figure*}
\includegraphics[width=.98\textwidth]{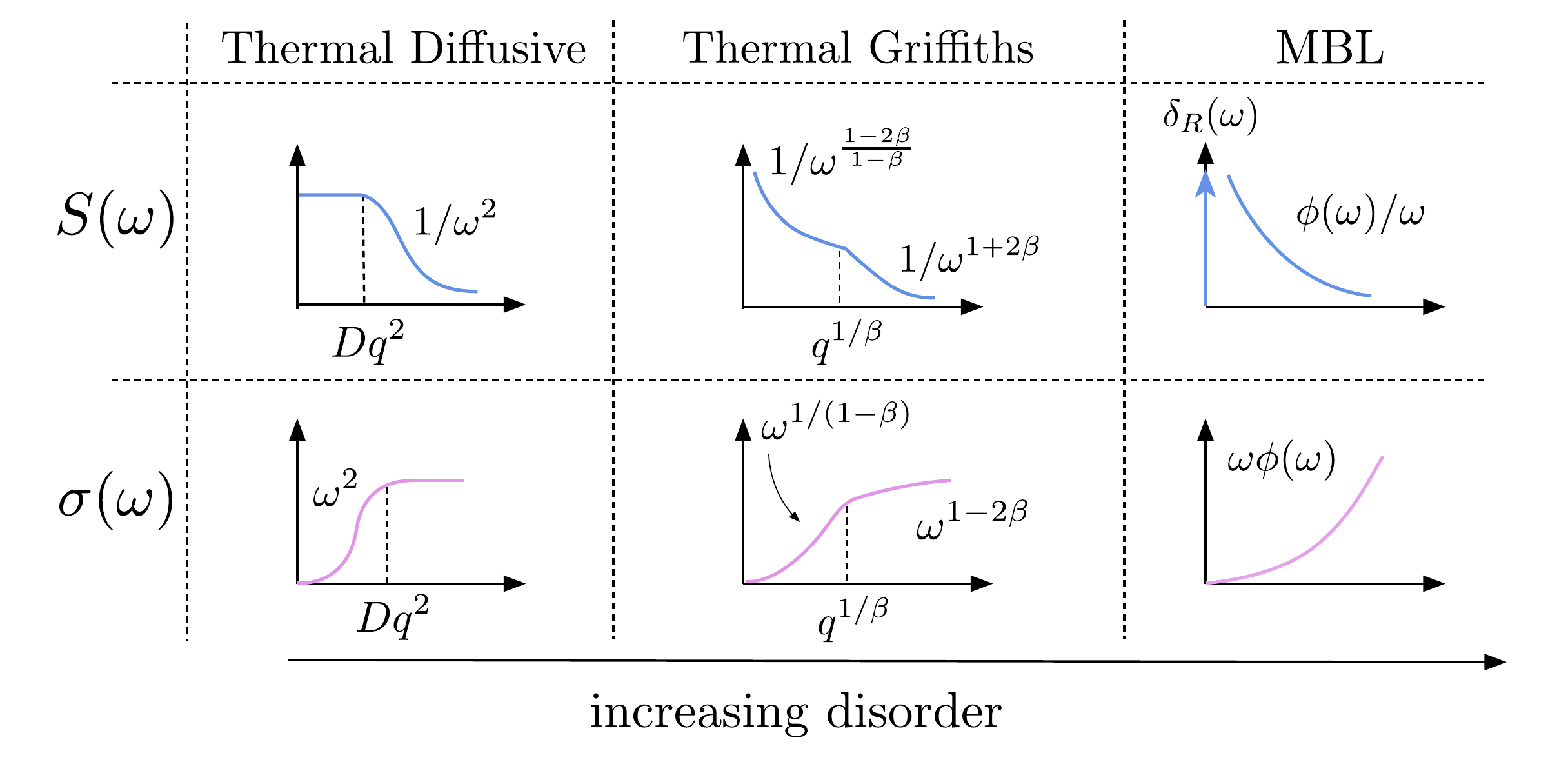}
\caption{Behavior of the frequency ($\omega$)- and wavevector ($q$)-dependent structure factor $S(q,\omega)$ and conductivity $\sigma(q,\omega)$ in disordered one-dimensional systems. Both $\omega$ and $q$ are assumed small compared with the characteristic microscopic scales of the problem. The response is schematically plotted as a function of $\omega$ at fixed nonzero $q$.
In the thermal phase (left and central panels) there is a crossover in the response when $\omega \sim q^{1/\beta}$, with $\beta = 1/2$ in the diffusive phase and $0 < \beta < 1/2$ in the subdiffusive Griffiths phase. In the MBL phase near the transition (right panel) the structure factor goes as $\phi(\omega)/\omega$ and the a.c. conductivity goes as $\omega \phi(\omega)$; the form of $\phi(\omega)$ is discussed in Sec.~\ref{sec:GriffithsinMBL}. The static $\sim\delta(\omega)$ contribution to $S(\omega)$ in the MBL phase is indicated.  There is no appreciable $q$-dependence in the MBL phase, except in logarithmic corrections.}
\label{responsediag}
\end{figure*}
\end{center}

\emph{Conductivity}. In the subdiffusive regime ($\beta<1/2$), the frequency- and wavevector-dependent conductivity $\sigma(q, \omega)$ has nontrivial structure as a function of $q$ and $\omega$~\cite{Agarwal,gadhk}; Fig.~\ref{responsediag}.  The behavior of the a.c. conductivity is an illustrative example of the interplay between various kinds of rare-region effects. Let us first consider the conductivity when $q \rightarrow 0$ at fixed $\omega$. In this case, the current is limited by $\omega$ rather than $q$, and the response is effectively $q$-independent.  On a timescale $1/\omega$, the system can be divided into segments of length $L \sim 1/\omega^{\beta}$ within which the system relaxes close to local equilibrium, but for $\beta<1/2$ these segments are separated by bottlenecks that do not allow relaxation on this time scale.  Suppose the system is driven with a spatially uniform force $E$ on the conserved particles that oscillates at frequency $\omega$.  Each internally equilibrating segment then relaxes to a density profile of the form $\rho(x) - \rho_0 \sim E x$, where $\rho_0$ is the equilibrium density at $E=0$.  The total displacement (from equilibrium) of the density distribution during the cycle is then $\sim \int dx x (E x) \sim EL^3$, so that the total current in this segment is $\sim EL^3 \omega$, and the current per particle is $\sim EL^2 \omega$. This leads to the relation that the low-frequency conductivity goes as 

\be
\sigma(q=0,\omega) \sim \omega^{1-2 \beta} = \omega^{(z-1)/(z+1)}.
\ee

Now let us consider the opposite limit of $\omega \rightarrow 0$ for finite $q$. In this case, the charge within a typical region of the system is able to rearrange completely before the field changes direction. The response of typical regions is therefore dominantly reactive. However, \emph{atypical} regions, i.e., inclusions that are insulating on timescales $1/\omega$, respond dissipatively at that timescale. The density of such slow inclusions goes as $\omega^{1/z}$, and each such inclusion has charge rearrangement at a rate $\omega$, so that the low-frequency behavior of the real part of the spatially-averaged conductivity at nonzero $q$ is 

\be
\sigma(q,\omega \rightarrow 0) \sim \omega^{(z + 1)/z} = \omega^{1/(1-\beta)}.
\ee

Note that the frequency-dependence in these two $q$-regimes is sharply different from the diffusive behavior of $\sigma(q, \omega) \sim \omega^2 / (\omega^2 + D^2 q^4)$.  Meanwhile, as the MBL phase transition is approached, $z\rightarrow\infty$ and both the low-$q$ and high-$q$ conductivities approach a $\sigma\sim\omega$ frequency-dependence.

\emph{Structure factor and relation to $1/f$ noise}.---The a.c. conductivity is closely related to the dynamic structure factor, 

\begin{align}
S(q, \omega) &\equiv \text{F.T.}[\langle [\hat{\rho}(q,t), \hat{\rho}(q,0)]_+ \rangle] \nonumber \\
&= q^2 \text{Re} [ \sigma(q, \omega) ] \coth{\left( \hbar \omega / 2 k_B T\right) }/ \omega ; 
\end{align}

Fig.~\ref{responsediag}. Recently, the $(q,\omega)$-dependence of the structure factor was explored numerically across the MBL transition~\cite{prelovsek2016a, prelovsek_dynamical_2016} as well as in bond-disordered spin systems with some analogies to MBL~\cite{adm}. We note this spectral function can be a ``noise'' spectrum: if an external qubit couples to the operator $O$ of the nearly MBL system---for instance, it has a Hamiltonian $H_q = \hat{\sigma}_z \epsilon_z /2 + \sigma_x O$---then the relaxation rate $1/T_1$ of the qubit, given by a sum of Fermi's Golden rule rates of absorption and emission of the qubit due to fluctuations of the operator $O$, is given by the analogous structure factor $S_O (\omega = \epsilon_z)$. In the Griffiths phase, for finite $q$, 

\be
S(q, \omega \rightarrow 0) \sim \omega^{-1 + 1/z}.
\ee

Thus, the transition into the MBL phase is characterized by the presence of spatially-averaged spectral function with a $1/\omega$ tail. Although this is an intriguing feature of the near-MBL dynamics, we caution that a number of different physical mechanisms can give rise to such low-frequency tails~\cite{1overfSolidsDuttaHorn}.

\emph{Griffiths effects beyond linear response: quench dynamics and heating}.---The arguments we have made to estimate rare-region contributions to linear response generalize directly to quench dynamics. In both cases, the crucial step in the reasoning is the assumption that rare insulating inclusions remember their initial configuration at late times. Thus, for example, our analysis of the large-$q$ conductivity and structure factor can be directly related to the decay of an imprinted density wave as measured in Ref.~\cite{Schreiber15}. Specifically, we expect that at a late time $t$, the remaining memory of the initial pattern is restricted to inclusions of size $\agt \zeta \log t$; the density of such inclusions, and therefore the residual contrast, decays as $t^{-1/z}$. 

One might wonder whether a more direct formal relation exists between linear response and the late-time dynamics after a quench. In some thermalizing systems, such a relation can be derived because the late-time density matrix of a system is close to equilibrium, thus, one can think of it as being perturbed slightly away from equilibrium, so that linear response theory applies. In the Griffiths picture, however, no such direct relation exists: deep inside inclusions that are insulating at time $t$ (and that, as discussed above, dominate response at time $t$), the system remains far from equilibrium, and is not in a linear-response regime. 

A complementary probe of Griffiths physics comes from \emph{nonlinear response}, captured, e.g., by the rate at which the system absorbs energy from an external drive of amplitude $A$ and frequency $\omega$~\cite{gkd}. We assume in what follows that the system is initialized at an energy density corresponding to finite temperature. In the thermal phase---ignoring Griffiths effects---energy absorption is initially linear, with a rate $\sim A^2$, and this linear response saturates on a timescale $W/A^2$ (on which the system has heated up). In the MBL phase, the system consists of essentially isolated two-level systems; again, the initial absorption is linear with a rate $\sim A^2$, but the average absorption of near-resonant two-level systems saturates on a timescale $\sim 1/A$~\cite{gkd}. A key distinction between these phases is that in the MBL phase, response is dominated by a small number of resonant transitions (so most degrees of freedom do not heat up at all, and the system enters a Floquet MBL phase, provided the drive is not too strong~\cite{bordia_periodically_2016, ldm, adh2014, gkd}) whereas in the thermal phase the entire system heats up. The response of a large MBL inclusion (with characteristic timescale $\tau \gg W/A^2$) in the thermal phase can be understood by a combination of these effects. The thermal region heats up and saturates (to infinite temperature) on a timescale $\sim W/A^2$, whereas the MBL inclusions direct absorption from the drive saturates on a timescale $\sim 1/A$.  On a timescale $\sim W/A^2$, then, the system has heated up to infinite temperature everywhere outside the inclusion, whereas the inclusion is still cold. Then the inclusion is thermalized by its surroundings on a timescale $\sim \tau$.  The energy absorption rate at late times $t$ thus goes as $1/t$ times the density of inclusions for which $\tau \agt t$, i.e., 
\beq
\frac{dE}{dt} \sim \frac{t^{-1/z}}{t} 
\eeq
so the energy approaches its final, infinite-temperature value as $E_\infty - E(t) \propto t^{-1/z}$. 

In addition to this long-time signature of Griffiths physics, one can detect the quick saturation of response inside inclusions via ``hole-burning'' experiments~\cite{black_halperin}, in which one first strongly drives a system at a particular frequency $\omega$, removes the drive and waits a time $T$, and then probes absorption at the same frequency $\omega$. The change in the spectral function $S(\omega)$ induced by the drive is then measured. The ``memory'' effect due to insulating inclusions should depend on the waiting time $T$ as $\Delta S(\omega) \sim T^{-1/z}$. This hole-burning signature will be clearest on intermediate timescales $1/A \ll T \ll W/A^2$, since at these intermediate times the thermal bulk of the system is still responsive to the drive. 

Finally, we remark that when the system is driven at amplitudes that are large compared with the drive frequency, the drive can cause some inclusions to delocalize~\cite{ldm, adh2014, gkd}. This effect can renormalize the exponent $z$, which should therefore be regarded in general as a function of the drive amplitude and frequency in the resulting thermal Floquet Griffiths regime.

\subsection{Broad distributions in the Griffiths phase}

We have seen that typical and spatially averaged quantities can differ strongly in the thermal Griffiths phase. This is related to the fact that such quantities have fat-tailed probability distributions. For quantities that relax locally (e.g., the high-$q$ conductivity or charge-density-wave relaxation), the relevant probability distributions can straightforwardly be read off from the distribution of inclusions that are insulating on timescale $t$: the probability of a given region having a characteristic relaxation time $\agt \tau$ goes as $\tau^{-1/z}$. 

In one-dimensional systems, for quantities where bottlenecks between typical regions dominate the dynamics, e.g., low-$q$ conductivity or entanglement growth, the argument is slightly different. One can map this system~\cite{Agarwal} to a resistor network with a resistance distribution $P(R) \sim R^{-\tau}$, where $\tau \equiv 1/(1 - \beta)$, where $\beta$ is the subdiffusion exponent. In the subdiffusive regime, $1 < \tau < 2$. The typical resistance of a sample of length $L$ is dominated by the largest resistor (slowest weak link) one expects to find in such a sample, which in turn is given by the requirement that $P(R > R_{max}) \simeq 1/L$. However, because of the power law distribution, an appreciable (i.e., only power-law small) fraction of samples contain a much larger resistance bottleneck. 

We observe that this broad distribution of physical observables also has implications for the distribution of matrix elements between eigenstates. The standard expectation from mesoscopic physics~\cite{imrydielectricanomalies,altshulerqplifetimefinite} is that, for a system of length $L$, eigenstates separated by less than the Thouless energy $D/L^2$ are effectively random.
This is also the content of the off-diagonal part of the eigenstate thermalization hypothesis.  However, the existing numerical evidence~\cite{serbynspectral,luitz2016anomalous} suggests that there is considerable structure (e.g., fat tails) in the distributions of these off-diagonal matrix elements in the subdiffusive thermal Griffiths regime. 
The precise implications of Griffiths effects for these eigenstate correlations remain to be worked out.

\subsection{Entanglement and operator dynamics}\label{thermal_ent}

In addition to ``bottlenecking'' the transport of conserved quantities, MBL inclusions slow down the growth of entanglement.  In clean systems, entanglement is expected to grow ``ballistically'', as the entanglement across a cut involves those degrees of freedom within a distance from the cut that grows linearly with time.  For concreteness, let us consider a one-dimensional system with open boundary conditions, initialized in a generic high-energy product state.  Thus, initially, its entanglement entropy across any cut is zero.  In a generic quantum-chaotic clean system the entanglement entropy grows linearly with time (with a ``velocity'' $v_E$), and saturates when the entanglement entropy of the smaller sub-region approaches its thermal value.  
The nature of this ballistic entanglement growth has recently been explored for generic strongly interacting systems~\cite{tsunami, mezei_stanford} and also for generic local systems subject to noise~\cite{nahum_haah}. 

In \emph{disordered} systems, it was initially argued \cite{VHA} that entanglement continues to grow ballistically so long as the dynamical Griffiths exponent (as defined above) $z < 1$.  The idea behind this was that once two large thermal regions are able to interact across a bottleneck, local interactions on either side of the bottleneck are able to fully entangle the two thermal regions. 
Currently the entanglement dynamics are not fully understood, but it seems clear that this early idea overestimated the rate at which entanglement spreads. A revised picture will be presented in Ref.~\cite{nahum_huse}. This revised picture implies that, whenever there is some nonzero density of bottlenecks with arbitrarily small rates, entanglement cannot grow with a sustained nonzero velocity, but always grows sub-ballistically. A brief sketch of the argument~\cite{nahum_huse} is as follows:
the inequalities derived in Ref.~\cite{nahum_haah} imply 
that the entanglement entropy across a cut at distance $L$ from a bottleneck exceeds the entanglement across the bottleneck by at most $\text{log}(q) L$, where $q$ is the local Hilbert space dimension.
Thus, for a given cut, if there is a bottleneck with rate $\Gamma$ at a distance $L$ from the cut, the entanglement across the cut is bounded by $S \leq \text{log}(q) L + \Gamma t$. It follows that if there is some nonzero density of bottlenecks with rate $\alt \Gamma_0$, the asymptotic entanglement velocity must be less than $\Gamma_0$. In the thermal Griffiths regime, there is a power-law distribution of rates over bottlenecks, $P(\Gamma < \Gamma_0) \sim \Gamma_0^{1/z}$ at small rates; thus, entanglement asymptotically grows as 

\be 
S(t) \sim t^{1/(1 + z)}, 
\ee

assuming that the inequality is saturated\footnote{In the simplest version of the Griffiths picture, which we have considered here, $z$ is the same for entanglement and transport. However, there might be refinements of the Griffiths scenario in which these exponents are different.}. 

\emph{Operator dynamics}.---Although fully entangling two thermal regions of size $L$ across a bottleneck requires one to in some sense move of order $L$ units of information across the cut, correlations can in principle spread faster \cite{nahum_haah, nahum_huse}.  For example, the hopping of a single particle across a weak link between two thermal regions of size $L$ can increase the end-to-end density correlations by of order $(1/L)$.  Thus, the spreading of correlations (and thus the ``light cone'') inside the Griffiths phase should be parametrically faster than the growth of entanglement.  Nevertheless, even the spread of correlations at time $t$ is limited by bottlenecks that are insulating on times $\agt t$. Thus, when we are near enough to the MBL transition so that $z>1$, the ``light-cone'' growth is also sub-ballistic, and over a time $t$ the distance over which correlations have spread is $\sim t^{1/z}$. It seems plausible that the rate of light-cone spread coincides with the rate of growth of the region in which out-of-time-ordered correlators are large; the latter have recently been explored by various authors in the MBL context~\cite{swingle_slow_2016,chen_universal_2016,huang_out--time-ordered_2016,fan_out--time-order_2016}. In Ref.~\cite{mezei_stanford} it is remarked that these rates are the same for clean chaotic systems, but it is not clear at present if they remain the same in strongly disordered systems.

\emph{Light-cone bound and the decay of typical autocorrelation functions}.---From the above arguments, we see that (provided the dynamical exponent $z > 1$) information that starts out at a particular point in the system can spread out over a distance $\alt t^{1/z}$ in a time $t$ (this is given by the light-cone bound above). A generic autocorrelation function is bounded from below by $1/N$, where $N$ is the size of the Hilbert space of the system. Thus, the \emph{most} a generic autocorrelation function can decay in time $t$ is given by $\exp(-\mathrm{const.} \times t^{1/z})$, i.e., as a stretched exponential in time. (Contrast this with the power-law decay of autocorrelations of conserved densities, as discussed above, which are bounded from below by the inverse volume of the system.) This would correspond to complete dephasing in the region over which correlations have spread. In systems with global conservation laws, generic correlation functions will not saturate this bound, instead decaying as power laws. Whether this bound is saturated in Floquet systems with no conservation laws, or whether a tighter bound (using the possibly slower entanglement velocity) can be derived in general, is currently an open question.

\subsection{Numerical evidence}

We now discuss some of the numerical evidence for the arguments presented above (see also Ref.~\cite{luitz2016ergodic} for a more detailed review). These numerical studies are based on studying one-dimensional models using various techniques, including exact diagonalization (ED) techniques and dynamic matrix product state (MPS) related methods.
The model that has primarily been investigated is the spin-1/2 XXZ chain with random magnetic fields in one fixed direction (usually denoted by $z$): $H = \sum_i h_i \hat{S}^z_i + \sum_{i} J \hat{S}^x_{i} \hat{S}^x_{i+1} + J \hat{S}^y_{i} \hat{S}^y_{i+1} + J_z \hat{S}^z_{i} \hat{S}^z_{i+1}$.
In $d = 1$, one may also think of this system as a model of fermions or hard-core bosons on a lattice~\cite{jordanwigner}, with the magnetic fields acting as local chemical potentials, in-plane ($x-y$) spin interactions corresponding to particle hopping, and out-of-plane ($z$) spin interaction corresponding to nearest-neighbor density-density interactions.  Following the early work by {\v Z}nidari{\v c}, et al.~\cite{znidaric_many-body_2008}, the magnetic fields are typically chosen to be distributed uniformly between $[-W,W]$ with $W$ characterizing the strength of the disorder.  The total magnetization in the $z$-direction (equivalent to the total particle number) is conserved and most studies choose to focus on the total spin sector $\sum_i \hat{S}^z_i = 0$ since the effect of interactions is maximized in this part of the Hilbert space.  This model has the special feature that when the random field is set to zero ($W=0$) it becomes an integrable system.  This gives the model nongeneric behavior for weak random fields. 
Diffusion in the clean-system limit can be restored by adding next-nearest neighbor interactions~\cite{moh}, at the cost of exacerbating finite-size effects.

Slow relaxation of autocorrelation functions was pointed out in Ref.~\cite{barlev_slow}; subsequently, subdiffusive behavior was noted by Bar Lev, {\it et al.}~\cite{BarLev} and Agarwal, {\it et al.}~\cite{Agarwal}. The latter work further presented evidence for the power-law frequency dependence of the conductivity and its broad distribution, and together with the early versions of Ref. \cite{VHA}, provided the interpretation of these results in terms of Griffiths physics.
These numerical studies, as well as subsequent work using exact diagonalization~\cite{luitz2016,kerala2015energy}, and approximate memory-matrix approaches~\cite{khaitmblmemorymatrix} found anomalous diffusion at essentially \emph{all} values of disorder. (In Ref.~\cite{Agarwal} an approximate value for the transition to diffusive behavior was estimated using finite-size scaling.) Subdiffusive behavior in the limit of weak disorder is unexpected from the point of view of Griffiths physics, as localized inclusions should become parametrically rare or altogether absent for sufficiently weak disorder.

A resolution to this issue was recently provided in the work of {\v Z}nidari{\v c}, {\it et al.}~\cite{znidaric_diffusive_2016}, who were able to explore much larger system sizes by an elegant application of matrix product operator techniques.  This work considers a finite XXZ spin chain, initially prepared in a fully mixed infinite temperature state, that is connected to simple baths at different magnetic fields at its two ends, such that in this system's nonequilibrium steady state there is (a) a spin density gradient and (b) a spin current along the chain. The dynamics of the full system plus bath can be described by a quantum master equation in Lindblad form~\cite{breuer_theory_2002, gardiner2004quantum}. Crucially, the steady state density matrix subject to the boundary driving terms is expected to be locally close to thermal so that it can be efficiently described as a low-rank matrix product operator, and can therefore be computed using a matrix product operator (MPO) based method~\cite{schollwock_density-matrix_2011, orus2014practical}.
This method yields both a diffusive regime at weak disorder and a subdiffusive regime at stronger disorder, as expected.  The boundary between these two regimes is at quite weak disorder compared to that at the MBL phase transition.  Could this small diffusive regime be a feature that is due to the system being integrable at zero disorder?  It will be interesting to see if this behavior is substantially different for models that are strongly nonintegrable at zero disorder.  Another interesting follow-up on this work will be to look at models where the field is quasiperiodic~\cite{iyerquasiperiodicmbl,naldesiaubreyandre} instead of random.  Nonrandom quasiperiodic models should not have Griffiths rare regions so may remain diffusive throughout the thermal phase, or at least to much closer to the MBL phase transition.

\section{Griffiths effects in the MBL phase}
\label{sec:GriffithsinMBL}

We now turn to the MBL side of the transition, and explore the dynamical effects of locally thermal inclusions. This topic is more subtle than that of MBL inclusions in the thermal phase: a locally thermal inclusion ``infects'' its MBL surroundings, thus thermalizing them, but the details and extent of this process are not fully understood.  
The earliest discussion of this topic was in a paper on \emph{classical} spin chains~\cite{oph}; it was argued that in the classical limit even small locally chaotic islands infect the entire system and thus inevitably restore transport.
As noted in Ref.~\cite{oph}, this argument does not directly apply to quantum systems: in these, the chaotic islands have a discrete level spacing, which in principle can ``cut off'' their ability to infect other degrees of freedom.
In one dimensional quantum systems with short-range interactions it is now clear that MBL is stable against such thermal inclusions at strong enough randomness~\cite{imbrieproof}. The situation in higher dimensional systems, or one-dimensional systems with longer-range interactions~\cite{burin_dephasing_1998, yao2013}, is not clear at present, as we shall see.

We will focus here on rare regions that are internally thermal; however, these are only one type of dynamically relevant rare region in the MBL phase. Because the dynamics in the MBL phase are exponentially sensitive to localization length, even inclusions that are slightly less (or more) localized than typical regions can dominate typical regions in response~\cite{mbmott}. These ``same-phase'' rare regions---which may be dominant deep inside the MBL phase---are outside the scope of this review.

\begin{figure}
\includegraphics[width=3.3in]{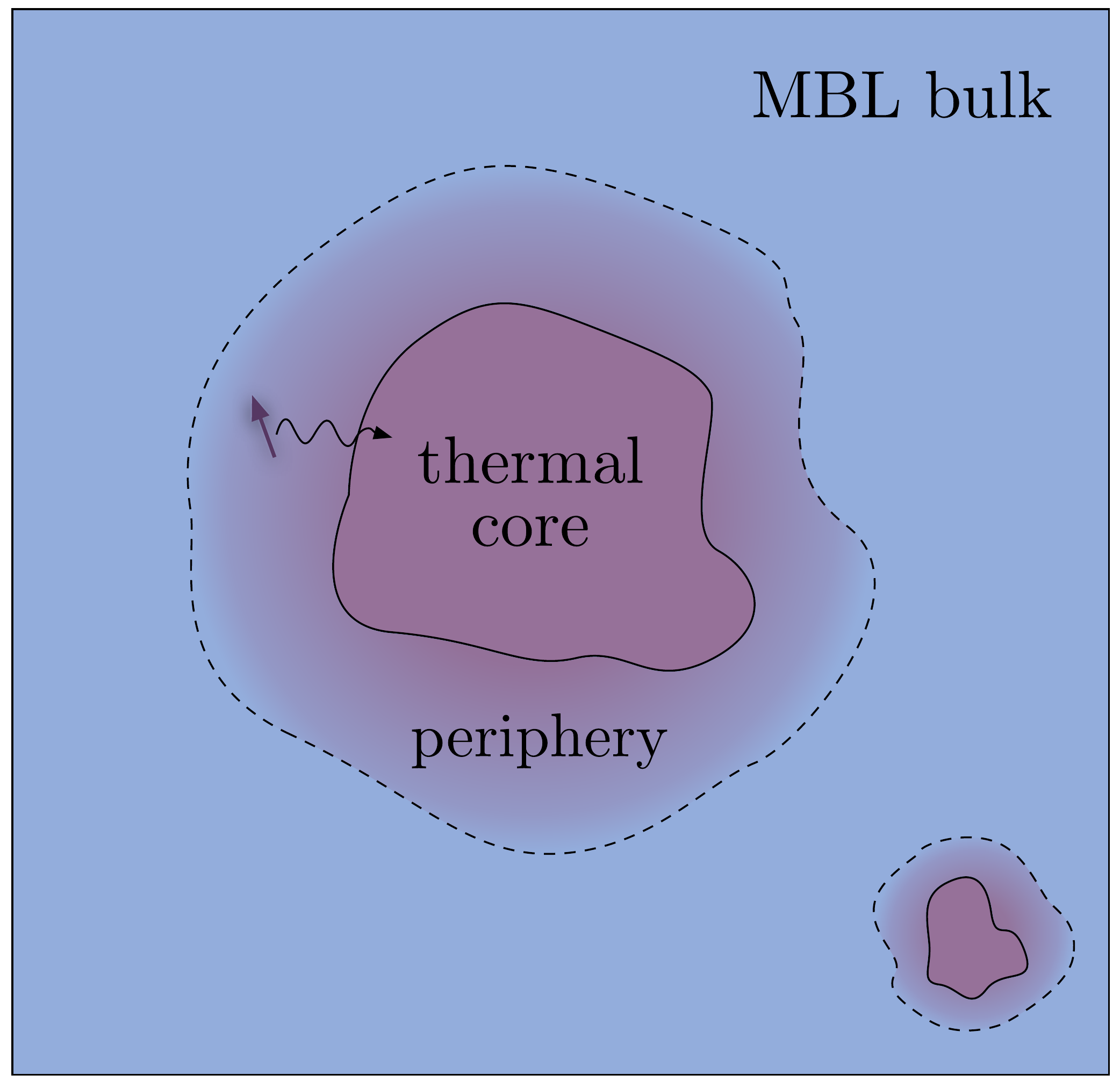}
\caption{A rare large thermal inclusion in the MBL phase.  The rare region is the central thermal core.  The periphery are the degrees of freedom that are well entangled with the core in the system's eigenstates.  The ``MBL bulk'' remains localized with only boundary-law eigenstate entanglement.}
\label{thermalcore}
\end{figure}

\subsection{Implications for response}

We now review the behavior of a single thermal inclusion in the MBL phase (Fig.~\ref{thermalcore}),
following the same logic as Sec.~\ref{STI}. Once again, the Hamiltonian can be separated into terms that act inside the inclusion (which extends to radius $R_0$), outside the inclusion, and across the boundary. The MBL region (now the \emph{outside} region) is described by the  Hamiltonian of local integrals of motion (or l-bits) $\hat H = \sum_i h_i \hat \tau^z_i + \sum_{ij} J_{ij} \hat \tau^z_i \hat \tau^z_j  + \ldots$. A maximally simple model for the thermal inclusion is as a random matrix; this corresponds to assuming that transport across the inclusion is effectively instantaneous. The boundary terms couple l-bits to the inclusion, with a coupling that decays exponentially with distance from the inclusion (and also decays, for processes that flip many l-bits that are not along a nearly straight path, with the number of l-bits involved). These couplings are, by construction, the only processes that can flip the l-bits. Anticipating that the thermal region will be able to thermalize some spins in the initially MBL region, we introduce the following terminology: the thermal ``core'' is the microscopically rare region, whereas the ``periphery'' consists of those initially MBL spins that are thermalized by the inclusion \cite{mbmott}.

The Golden Rule analysis for a finite-sized bath to thermalize a single two-level system proceeds as follows~\cite{MBLBath}:  Let us consider the simplest l-bit/bath coupling, which is of the form $g \hat \tau^x_i \hat O_i$, where $\hat \tau^x$ flips a particular l-bit. The operator $\hat O_i$ has matrix elements $\sim 2^{-R_0^d/2}$ between any pair of states, where $R_0^d$ is the number of spin-1/2's in the inclusion; we have absorbed geometric factors of order one in to our chosen units of length.  For simplicity, we choose the energy of the thermal inclusion and the periphery to correspond to infinite temperature, but the results do not change qualitatively for nonzero finite temperatures.  Thus the l-bit flip strongly mixes bath levels whenever $g 2^{-R_0^d/2} \agt W  2^{-R_0^d}$, where $W$ is the bandwidth of the inclusion.  However, $g \sim \exp{(-R/\zeta)}$ typically falls off exponentially with the distance $R$ between the l-bit and the thermal inclusion. Thus, at this elementary level of analysis where we treat each l-bit independently, we expect that whenever $\exp(-R/\zeta) \agt 2^{-R_0^d/2}$, that l-bit will be thermalized by the bath. It was later argued in Ref. \cite{drh} that the peripheral spins thermalized by the bath are absorbed in it and make the bath stronger by reducing its level spacing. An approximate analysis of this renormalization led to the conclusion that a single thermal inclusion can destabilize the entire MBL state in two dimensions or higher. We will discuss this scenario later.  

We now discuss the effect of thermal inclusions on the dynamic response in the MBL state assuming, for  now, that the inclusions only generate a finite thermal periphery, i.e. that the MBL phase is stable to large inclusions.  
We will also assume that the inclusion is spatially compact (we will visit the question of fractal inclusions below) and described by random-matrix theory.  As noted earlier, the thermal ``core'' of the inclusion is assumed to relax rapidly, and thus does not contribute to conductivity or dissipative response at low frequencies (or, equivalently, to dynamics at long times). However, the core of the inclusion is surrounded by a ``periphery'' of l-bits that are incorporated into the inclusion as discussed above.  These peripheral l-bits have a broad distribution of relaxation rates.  The slowest relaxation rate is that of the l-bits at the farthest edges of the periphery, which relax at a rate that is exponentially small in the total volume of the core and periphery.
When probed at a low frequency $\omega$, response will be dominated by inclusions for which these slowest l-bits relax at rate $\sim\omega$,
as these are more common than larger inclusions.  For these inclusions the width of the periphery is only $\sim\log\omega$ or less.

The precise nature of this low-frequency contribution depends on the response function being probed. For the specific case of the low-frequency conductivity, we note that the decay of a peripheral l-bit at rate $\omega$ involves moving a charge a distance (ignoring factors of $\log{\omega}$) of order unity at a rate $\omega$. Therefore the conductivity varies as $\omega$ times the density of contributing inclusions. Other response functions, such as the structure factor, go as $1/\omega$ for a single inclusion. The density of relevant 
inclusions is exponentially small in the volume of their core.  First, we will consider inclusions with a compact core and assume that the ratio of the volume of the periphery to the volume of the core remains finite for large inclusions.  In this case the relevant inclusions are then
polynomially rare in $\omega$, i.e., $n(\omega) \sim \omega^g$; therefore, this analysis of compact cores suggests that low-frequency spectral functions vanish (or diverge) with continuously varying powers of $\omega$ on the MBL side of the transition.  The Griffiths exponent $g>0$ is expected to vanish as we approach the MBL transition and grow large as we go ``deep'' in the MBL phase far from the transition.  Specifically, the conductivity would go as $\sigma(\omega) \sim \omega^{1 + g}$ near the MBL transition, while the spectral function would go as $S(\omega) \sim \omega^{g - 1}$. This Griffiths power-law competes with a separate continuously varying power-law due to many-body resonances~\cite{mbmott}, which may dominate it deep in the MBL phase. (Note, however, that generic spectral functions such as the structure factor \emph{also} have a zero-frequency ``Drude'' contribution coming from typical regions.)

Finally we note that the $q$-dependence of the structure factor or conductivity in the MBL phase is weaker than in the thermal phase. We have so far been implicitly considering the response at wavelengths that are long compared to the size of an inclusion. However, even for shorter-wavelength modulations (i.e., $q \gg \zeta \log (W/\omega)$), the $\omega$-dependence of the conductivity will remain unchanged up to logarithmic corrections due to changes in the relevant dipole matrix elements.

\begin{center}
\begin{figure}
\includegraphics[width=3.3in]{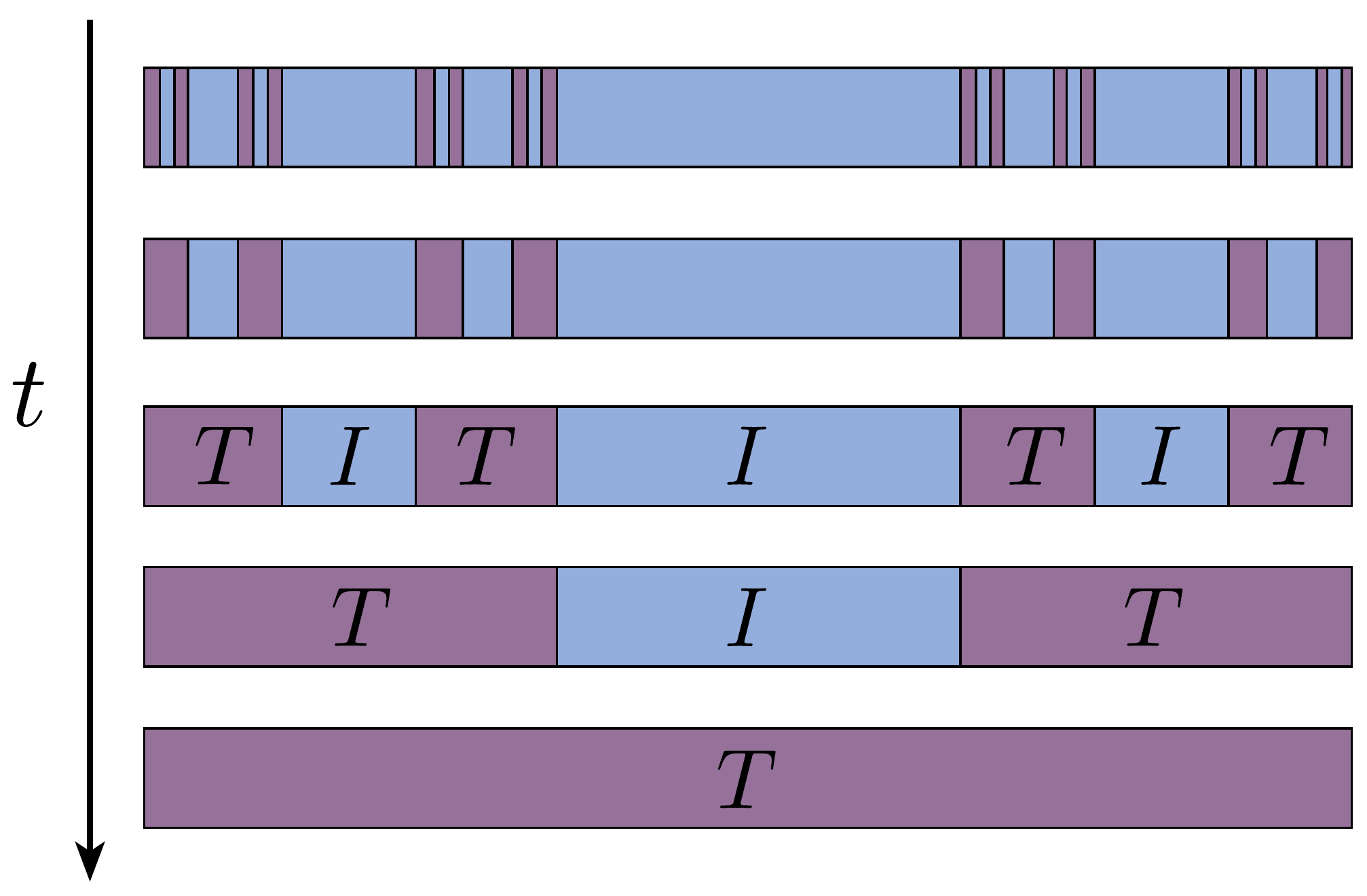}
\caption{Illustration of the possible fractal nature of thermal inclusions in one dimension~\cite{ZhangTITMoves}.}
\label{fractalfig}
\end{figure}
\end{center}

\subsection{Fractal inclusions}

When one models an inclusion by random-matrix theory, one implicitly throws out information about its spatial structure. Spatial structure only enters through the fall-off of l-bit couplings. This observation means that, \emph{provided} the random-matrix picture is appropriate for fractal inclusions, these are in fact the dominant source of low-frequency response. The reasoning is as follows~\cite{ZhangTITMoves}:  Consider a system that contains two thermal (T) regions that are separated by a sufficiently thin insulating (I) typical region (which is a small enough fraction of the size of the thermal regions). Evidently the insulating region will be thermalized.  Now we can iterate this argument, supposing that each thermal segment itself has the same ``TIT'' structure, as illustrated in Fig.~\ref{fractalfig}.  The endpoint of this argument is that the most probable thermal inclusion of a certain size is one in which the microscopically ``thermal'' (and therefore rare) core is a vanishing fraction, occupying a fractal network of fractal dimension $d_f<d$.  If the Golden-Rule arguments above apply to such inclusions, it follows that the probability cost of an inclusion goes only as $p^{L^{d_f}}$, where $L$ is the linear size of the inclusion; in one dimension the density of such inclusions thus decreases \emph{slower} than a power law of their relaxation frequency $\omega$.  This then effectively sets the Griffiths exponent to $g=0$ (up to logarithms).
This conclusion (like that in Sec.~\ref{singleETHinclusion}) follows naturally from the random-matrix assumption, but the assumption itself (and the associated neglect of spatial information) appears particularly questionable for a fractal inclusion.

We briefly discuss the implications of this ``fractal-inclusion'' scenario for response. A generic spectral function in the MBL phase (i.e., one that has a Drude peak, such as the structure factor) would then go as $\sim(1/\omega) \exp[-c (\log (W/\omega))^{d_f}]$.  Meanwhile, the low-frequency conductivity in the MBL phase goes as $\sigma(\omega) \sim \omega \exp[-c (\log (W/\omega))^{d_f}]$.  We note that (depending on $d_f$ and the available dynamical range) this dependence can be hard to distinguish from a power law in small-system numerics.

\subsection{Potential instability of  MBL to thermal inclusions}\label{singleETHinclusion}

We now revisit the fate of a single thermal inclusion in an MBL phase without assuming that the periphery around it is finite. We will review the arguments used in Ref. ~\cite{drh} that lead to the conclusion that MBL is unstable in dimensions $d>1$. We emphasize that this conclusion rests on an approximate treatment and we indeed raise several possible critiques of the analysis, at the end of this section, that may invalidate its conclusion.  

We first examine how the bath changes by thermalizing the first peripheral spin. Let this spin have a Hamiltonian $H_s = \omega_1 \hat \tau_1^z$. Before the boundary couplings are included, we can describe the $2^{R_0^d + 1}$ energy levels of the core plus this one peripheral spin as product states: $|\psi\rangle_0 = |\pm\rangle_{\mathrm{l-bit}} |n\rangle_{\mathrm{core}}$. In what follows we shall drop the subscripts. When the boundary couplings are included, states (of the full system of $R_0^d + 1$ spins) that are resonant to within the l-bit's Golden Rule linewidth $\sim g^2/W$ hybridize; note we are assuming the l-bit-flip matrix element is large compared to the many-body level spacing of the inclusion. Then the eigenstates at energy $|E\rangle$ can be written in the  form:
\beq\label{huveneers1}
|E\rangle = |+ \rangle \sum_{n_+ } A_{n_+} |n_+\rangle + | - \rangle \sum_{n_-} A_{n_-} |n_-\rangle ~,
\eeq
where the sums over $n_\pm$ runs over (unperturbed) eigenstates of the thermal inclusion.  The amplitudes $A_{n_{\pm}}$ are significantly nonzero for $|E - E_{n_{\pm}} \mp \omega_1 | \alt g^2/W$ and are effectively random in sign and magnitude and are of the order to properly normalize this state.  The total weight in this state is almost exactly equally divided between the l-bit being up and being down.  Thus this l-bit is well thermalized by the inclusion. We note again that these arguments apply only when the line-width $g^2/W$ is much larger than the many-body level spacing of the inclusion.

We can now discuss the generic features of the spectral function of a different operator $\hat O$ that acts only on the original thermal core (that we will later couple to a different peripheral l-bit).  This involves computing matrix elements of the form $\langle E | \hat O | E' \rangle$ (where $|E'\rangle$ also has the form~\eqref{huveneers1}).  Note that $\hat O$ does not act directly on the first peripheral spin, so only terms that are diagonal in the $|\pm\rangle$ l-bit basis of Eq.~\eqref{huveneers1} contribute:
\beq\label{huveneers2}
\langle E | \hat O | E' \rangle = \sum_{\pm} \sum_{m_{\pm},n_{\pm}} A^*_{m_{\pm}}A'_{n_{\pm}} \langle m_{\pm}| \hat O |n_{\pm}\rangle ~.
\eeq 
We now make the superficially reasonable assumption that for a large inclusion the amplitudes and matrix elements that are in this sum~\eqref{huveneers2} are effectively uncorrelated random variables\footnote{When the spectrum of $\hat O$ is being probed at frequencies $\omega \alt g^2/W$, there will be terms in the sum~\eqref{huveneers2} for which $|m_\pm\rangle$ and $|n_\pm\rangle$ coincide. These terms involve \emph{diagonal} rather than off-diagonal matrix elements of $O$. The diagonal matrix elements have a mean value given by ETH, and their fluctuations about their mean value are uncorrelated random variables on the same scale as the typical off-diagonal matrix element. One can explicitly check that the mean values cancel in $\overline{|\langle E | \hat O | E' \rangle|^2}$, leaving only fluctuations. Thus, at this order of analysis, the low-frequency and high-frequency behavior appear identical.}.  
Then when we evaluate the mean-square matrix element we obtain $\overline{|\langle E | \hat O | E' \rangle|^2}=\overline{|\langle m_{\pm}| \hat O |n_{\pm}\rangle|^2}/2$.  Thus the first peripheral l-bit getting entangled with and thus ``joining'' the thermal inclusion has reduced the inclusion's many-body level spacing by a factor of 2, but has not affected the Golden Rule linewidth of other peripheral l-bits coupled to the inclusion.
This is in some ways a surprising result, and is not strictly valid, as pointed out in Ref.~\cite{drh}, although the corrections they consider (due to the ``back-action'' of the peripheral spin on the bath) turn out to be small. A rough intuitive justification for the result might run as follows: at short times (compared with the decay rate of the first peripheral spin) the bath does not feel the peripheral spin, but the Heisenberg ($\sim 1/t$) broadening of the bath spectral lines is large enough that the bath looks effectively continuous anyway.  On long timescales compared with the decay time of the first peripheral spin, the bath and peripheral spin are fully entangled, so it is appropriate to think of the effective size of the bath as having increased.

De Roeck and Huveneers~\cite{drh} assume that the spectral function inside the \emph{initial} core remains featureless but increasingly finely spaced as more spins are absorbed. We shall revisit this assumption below; first, however, we discuss its implications in $d$-dimensional systems.  At each distance $R$ from the boundary of the inclusion core, one must compare the matrix element for flipping an l-bit---given by $\exp(-R/\zeta)/2^{c_d(R_0 + R)^d /2}$, with the renormalized level spacing of the inclusion plus all closer l-bits, $\sim 2^{-c_d(R_0 + R)^d}$, where we are assuming the density of l-bits is one, and the factor $c_d$ relates the volume to the radius of the sphere, e.g. $c_1=2$, $c_2=\pi$, etc.  In one dimension when $\zeta < 1/\ln 2$, the matrix element falls off faster with distance, and l-bits far enough from the inclusion are not thermalized.  The boundary of the peripheral region is then given by $R_1 = R_0\zeta\ln{2}/(1 - \zeta\ln 2)$.
For $d > 1$, on the other hand, (or in one dimension with slower falloff of the interactions) the level spacing is asymptotically smaller, so that a sufficiently large thermal inclusion appears, by this simple argument, sufficient to destabilize the MBL phase.  However, even if the MBL phase is unstable to thermalization by such rare inclusions, the time scale for this to happen may be so inaccessibly large that such systems act many-body localized at all accessible time scales.

The argument sketched above relies heavily on the assumption that the inclusion can be modeled as a random matrix, and that it remains random-matrix-like even after ``absorbing'' l-bits that it thermalizes. Thus there are various ways that this argument could fail. One possibility is that a pure random matrix might simply be an unrealistic model of a thermal inclusion, as it neglects any internal structure of the inclusion.  For instance, it cannot be true that all the matrix elements within a finite energy window are uncorrelated random numbers: there are $\sim 2^N$ such eigenstates, but the inclusion's Hamiltonian is typically specified by only $\sim N$ random numbers. These correlations are neglected in the random-matrix approach, but they might change the spectral structure of the inclusion in qualitative ways. Second, even if the random matrix assumption is initially valid (so that the early stages of the iterative construction are valid), it might eventually break down in the limit when the number of absorbed degrees of freedom greatly exceeds the number of degrees of freedom in the bare inclusion.
A third possibility is that although the spectral functions retain a relatively ``flat'' form as l-bits are added, the Golden Rule ceases to apply, because the bath spectrum has nontrivial higher-order correlations. This could arise, e.g., because of ``spectral diffusion''~\cite{nandkishore2016general}, where a relatively broad spectral lineshape comes about due to the slow spectral wandering of narrow lines. In the presence of spectral diffusion, the ``true'' correlation time of the bath is much longer than what one would get by Fourier transforming its spectrum.
At the moment, however, it is unclear whether any of these arguments actually applies. We emphasize that there is no compelling evidence that the MBL phase can be fully stable in higher dimensions, and the numerical evidence on one-dimensional systems with power-law or stretched-exponential interactions is also inconclusive.

Although, as argued above and in~\cite{burin_dephasing_1998, yao2013, hauke_heyl}, power-law long-range interactions can destabilize MBL, there is a MBL-like phase in the infinite-range quantum random energy model (QREM)~\cite{lps,blps}.  The relation (if any) between this MBL-like phase in the QREM and MBL phases with finite-range interactions is still not clear.

\subsection{Rare regions of the \emph{state}, many-body mobility edges}

The rare regions we have been considering so far are rare regions of the quenched disorder, i.e., regions in which all relevant eigenstates are in the same phase. It has also been proposed (famously by Basko, {\it et al.}~\cite{BAA}) that there is a temperature-tuned (or, more precisely, energy-density tuned) transition between thermal and localized eigenstates. This scenario is often called a ``many-body mobility edge.''  A many-body mobility edge differs from a \emph{single-particle} mobility edge in that the former occurs at extensive energies (nonzero excitation energy densities) whereas the latter occurs at intensive energies. It is clear that many-body mobility edges are incompatible with the l-bit phenomenology, as the putative thermal states in the middle of the spectrum are evidently not constrained by infinitely many local conservation laws.  In fact, at present there is no clear phenomenology of such partially MBL states.  However, it is intuitively clear that such many-body mobility edges must exist as sharp crossovers if not as true phase transitions, as most simple criteria for the many-body (de)localization transition---e.g., the proliferation of many-body resonances---naturally involve a factor of the entropy density.  

Once again, our focus is on states that are ``globally'' in the MBL regime (their total energy puts them on the MBL side of the putative mobility edge, which for specificity we take to be at low energies, i.e., ``below'' the mobility edge). Can such states have local energy density fluctuations that take them \emph{locally} across the mobility edge? It is clear that \emph{static} fluctuations of this type cannot exist in a many-body eigenstate: i.e., in an eigenstate the expectation value of the energy density in \emph{any} region has to be at or below the mobility edge. (If it were not, one could draw a slightly larger region that would still locally be on the thermal side of the mobility edge, in which case the energy density would thermalize and thus be uniform over the slightly larger region, and so on, until the fluctuation had been ``spread out'' to the point where there are no static energy fluctuations across the mobility edge.) However, \emph{dynamic} fluctuations are still possible: the energy density in a region is not a conserved quantity, so it is consistent for a particular measurement of the instantaneous local energy density to give a result that is above the mobility edge, so long as the \emph{time-averaged} local energy density is not. If the state is globally MBL, correlations of the energy density fall off exponentially on scales beyond the localization length. Thus, the presence of transient anomalously high-energy regions in a globally MBL eigenstate would have to be correlated with that of nearby anomalously low-energy transients. 

Keeping this general stability constraint in mind, let us turn to the arguments of Ref.~\cite{drhms}. This work considers, as a starting point, a system consisting of mesoscopic ``grains''. In the absence of intergrain couplings, the energy density of each grain is conserved. It is assumed that the high-energy (or ``hot'') states on each grain are ``thermal'' (i.e., random-matrix-like) whereas the low-energy (``cold'') states are ``localized''. The eigenstates of this decoupled-grain problem form a basis for the coupled-grain problem. Ref.~\cite{drhms} begins with a particular basis state (consisting of a sufficiently long sequence of hot grains, in an otherwise cold background) and argues that even for weak inter-grain coupling, the original basis state is unstable because it is resonant with another basis state containing a slightly longer sequence of slightly less hot grains. 
Note that the same resonant process can \emph{also} be constructed starting from a basis state in which all the hot grains are translated by one site.
Assuming that this ``spreading'' is the dominant resonant process, Ref.~\cite{drhms} concludes that a resonant, transport-enabling network exists, whereby a sequence of hot grains can move in an ``inchworm''-like way, by repeatedly expanding and then contracting with a net translation. 

As noted above, the constraint on the absence of \emph{static} thermal regions already implies that hot grains in the initial, uncoupled-grain problem must form resonances of some kind (in agreement with Ref.~\cite{drhms}). Thus, the instability of hot grains is not in itself a surprise (although it might well be that Ref.~\cite{drhms} substantially overstates the associated rates). Moreover, if we expand an uncoupled-grain eigenstate containing a thermal bubble in terms of exact eigenstates of the weakly-coupled-grain problem, essentially all the weight in this superposition must come from exact eigenstates with local energy density at or below the putative many-body mobility edge. (Weak inter-grain couplings affect the energy density only weakly.) Thus there must be \emph{some} eigenstates below the mobility edge that have appreciable weight on configurations containing hot bubbles, i.e., it cannot be held that bubble configurations are mobile only because they have admixture from above the many-body mobility edge. We are left with the following possibilities:

(i)~Typical low-energy eigenstates do in fact delocalize along the resonant network, so that there are no true many-body mobility edges, only crossovers.  This is the simplest conclusion and is that drawn by Ref.~\cite{drhms}.

(ii)~Mobile eigenstates do exist that have weight on the resonant network, but at low energy densities there are also eigenstates that avoid all ``hot bubble'' configurations and therefore remain localized.  This scenario in which localized and delocalized states coexist at the same energies was ruled out for single-particle localization by Mott, but Mott's reasoning might need modification in the many-body context.  Even if this were true, however, it would not count as a ``mobility edge'' scenario, since it implies that delocalized states exist down to the lowest energies.  But it would be a breakdown of the ETH in that some of the low-energy-density eigenstates are not thermal.

(iii)~Effects beyond the simple resonant-network construction restore localization.
Sometimes (as in random-hopping models, or hopping models with binary disorder~\cite{kramer_mackinnon}) naive resonance-counting incorrectly suggests delocalization. In these problems, some of the resonances identified by resonance-counting methods are pre-empted by even stronger resonances. For instance, in the hopping model with binary disorder, faraway initially resonant sites acquire self-energy corrections that detune them from resonance with one another. It is possible that some analogous effect can successfully pre-empt the spreading of localized bubbles\footnote{A specific possibility is that transient bubbles arise as a result of rare coincidences of phase~\cite{rigol_thermalization_2008} amongst independent resonances, and that they ``fragment'' (dissipating their energy into a larger region) faster than they can spread hydrodynamically.}.  However, it is not clear at present if any such larger-scale effect does succeed in localizing hot bubbles. 

To summarize, there is at present no consensus on whether these rare regions of the state in fact preclude many-body mobility edges. On the one hand, a mobility edge is clearly seen in all extant numerical studies~\cite{lfa2015}, but this could well be ``only'' a crossover. 
On the other hand, the theoretical construction of Refs.~\cite{drhms} is in some sense ``low-order'' and thus not conclusive. 
Thus, as with many other open questions concerning MBL, it seems the question of many-body mobility edges requires further technical advances to be cleanly answered.

\subsection{Static properties of the MBL Griffiths phase}

In this section we briefly summarize various further implications of the Griffiths effects discussed above for static quantities in the MBL phase. For specificity (and to avoid issues related to the potential instability of MBL in higher-dimensional systems, discussed above) we restrict our discussion to one dimension. 

\emph{Static correlations}.---Above, we discussed how rare regions can dominate dynamical response. We now briefly turn to their potential impact on \emph{static}, or equal-time correlation functions in eigenstates of the MBL phase. We focus on systems with a global conservation law, and consider the behavior of the associated equal-time density-density correlation function. In a typical region of the MBL system, this decays exponentially on a scale $\zeta$, whereas, in a thermal region of size $L$, density-density correlations decay as $1/L$~\cite{palmbl}. The spatially averaged density-density correlator at large distances $R$ then behaves as $\exp(-R/\zeta) + n_L/L$, where $n_L$ is the density of inclusions of size $\agt L$. Compact inclusions of length $L$ are exponentially rare in $L$, thus do not change the parametric behavior of these static correlation functions (although they might cause the typical and average localization lengths to differ). However, for fractal inclusions, $n_L \sim \exp(-\mathrm{const.} \times L^{d_f})$, where $d_f$ is the fractal dimension. Therefore, disorder-averaged equal-time density-density correlations should decay as stretched exponentials in this case. 

\emph{Distribution of l-bit couplings}.---In addition, the ``l-bits'' inside a thermal inclusion will be spread out over the entire inclusion. A simple estimate for an inclusion of size $L$ is that the interaction between any two l-bits in the inclusion is of order $\exp(-sL)$, where $s$ is the entropy density of the inclusion. Deep in the MBL phase we can assume that $s \zeta \ll 1$, in which case the coupling between l-bits inside an inclusion of size $L$ greatly exceeds the coupling between typical l-bits at distance $L$. In this regime, the probability density for two l-bits at a distance $L$ to have a coupling $\exp(-sL) \alt J \alt \exp(-L/\zeta)$ can be seen to go as $P(J) \sim J^{-1} \exp[-\mathrm{const.} \times (\log W/J)^{d_f}]$. There is suggestive recent numerical evidence that the couplings in the l-bit Hamiltonian are indeed broadly distributed~\cite{pcor}, with the couplings following the form $P(J) \sim 1/J$ deep in the MBL phase.

\emph{Eigenstate entanglement}.---The reasoning above also has implications for the distribution of entanglement entropy~\cite{bauerarealaw} for regions of size $L'$ in the MBL phase. The tail of this distribution is sensitive to rare-region effects. For instance, if $\zeta \ll L \ll L'$, the probability of getting entanglement $\agt L$ should correspond to the density of thermal inclusions of size $\agt L$. If the dominant thermal inclusions are compact, this density should be exponentially suppressed in $L$; however, if the dominant inclusions are fractal, there should be a stretched-exponential tail to the entanglement entropy distribution. 

\subsection{Numerical evidence}

A detailed exploration of the power laws seen in conductivity near the MBL transition was presented in Ref.~\cite{mbmott}. In that work, the response of a \emph{single} Griffiths inclusion was also explored, by choosing the first four sites of a twelve-site spin chain to have weak disorder, thus ``imprinting'' a thermal inclusion. Because the length-scales associated with $\sigma(\omega)$ only diverge logarithmically with $\omega$, it is possible to go to quite low frequencies before finite-size effects become significant~\cite{mbmott}. Some numerical work has also been done (in Ref.~\cite{drh}) on a model of single spins coupled to a random matrix with progressively weaker couplings; this work suggests that an inclusion that is initially a pure random matrix does become a better bath as it progressively absorbs spins, consistent with the theory set out in Ref.~\cite{drh}. However, the size of the absorbed ``periphery'' was smaller than or comparable to the size of the random-matrix core, so the asymptotic behavior of a small core that has absorbed a very large periphery is still untested. 

\emph{State-dependence}.---Recent diagonalization studies~\cite{klsh,EEdist_Xiongjie,longtail_luitz} of eigenstate entanglement near the MBL transition found strong state-to-state fluctuations of the half-chain entanglement entropy.  The presence of these fluctuations suggests that our picture so far, in which a particular region is either thermal or localized, is not an accurate description of the physics for the rather small systems that can be numerically diagonalized. 
An interesting conclusion of Ref.~\cite{klsh}, supported by their numerical data, is that the dominant entangling inclusions near the MBL transition are ``sparse''~\cite{PVP} and unable to fully thermalize their surroundings, and they are state-dependent.  This is rather distinct from the picture developed in the preceding discussion, but the regime being studied is more the finite-size critical regime rather than the Griffiths regimes of the two phases.
It is also unclear to what extent the results in Ref.~\cite{klsh} extend to large systems (at the system sizes that paper studies, the correlation length is clearly not in its asymptotic regime~\cite{chandran2015finite}). 

\section{Relation to theories of the MBL transition\label{sec:MBLtrans}}

In the previous sections we have seen that response on both sides of the MBL transition is dominated by regions that are locally in the other phase. This picture of the near-transition dynamics suggests that the transition itself may have the following structure: at each length (or time) scale the critical state is highly inhomogeneous, consisting of thermal and localized regions. As one coarse-grains, certain formerly insulating regions become thermal (because they are ``infected'' by nearby thermal regions) and some formerly thermal regions become insulating (because their many-body level spacing becomes large compared with the running frequency scale). In the asymptotically thermal (MBL) phase, coarse-graining makes the system increasingly thermal (localized); at the critical point, the distribution of thermal and localized regions is self-similar under coarse-graining (but with the localized regions predominating~\cite{VHA,PVP}). Thus, the construction of the MBL critical point is similar to an \emph{iterated} version of the Griffiths analyses we have gone through here. This is the basic idea underlying the strong-randomness renormalization-group treatments of the MBL transition~\cite{VHA,PVP,ZhangTITMoves}. 

\section{Concluding Remarks}

Here we have reviewed what is currently believed about the influence of rare regions on dynamics near the many-body localization transition, on both sides of the transition. 
The qualitative behavior of transport on the thermal side of the transition at least is mostly settled: both Griffiths estimates~\cite{Agarwal,VHA} and large-system numerical approaches~\cite{znidaric_diffusive_2016} indicate that there is a diffusive regime at weak disorder as well as a subdiffusive regime at stronger disorder. 
However, various questions, concerning the growth of entanglement and the \emph{typical} behavior of correlation functions in one dimension, are still somewhat poorly understood.  
Moreover, although we expect no subdiffusive regime in higher-dimensional systems (see however, Ref.~\cite{barlevslow2d}), in one-dimensional systems with interactions that fall off sub-exponentially with distance, or systems with quasiperiodic potentials, there is no clear numerical evidence either way for such systems. 

The effect of rare thermal regions in the MBL phase is, by contrast, much more poorly understood. The central open problem is understanding the effects of a single thermal inclusion in the MBL phase as a function of its size and other properties.  At the simplest level of analysis (in which a thermal inclusion is treated as a random-matrix, and its influence on the MBL bulk is described in terms of Golden Rule estimates~\cite{drh}) it seems that the effects of a single thermal inclusion can be so far-reaching as to preclude a stable MBL phase in systems in higher than one dimension, in the presence of quenched randomness. However, this analysis---in common with works that map the MBL problem onto a Bethe lattice hopping problem---ignores all correlations between matrix elements, and it is possible that such correlation effects could ultimately restore localization. If, however, this nonperturbative instability really exists, it seems that various aspects of MBL phenomenology, such as the possibility of localization in higher-dimensions or with long-range interactions, or the existence of continuous thermally averaged local spectral functions in the MBL phase (the so-called ``weak MBL'' regime~\cite{MBLBath}) can only be realized in systems without quenched randomness, such as quasiperiodic systems. 

\begin{acknowledgments}
We thank  D. Abanin, R.N. Bhatt, I. Bloch, P. Bordia, A. Chandran, W. De Roeck, F. Huveneers, V. Khemani, C. Laumann, M. Lukin, I. Martin, M. M\"uller, A. Nahum, R. Nandkishore, V. Oganesyan, A. Pal, F. Pollmann, G. Refael, U. Schneider, R. Vasseur, R. Vosk, N.Y. Yao, and many others for interesting discussions and collaborations. KA acknowledges support from DOEBES Grant No. DE-SC0002140. EA acknowledges support from ERC synergy grant UQUAM. ED acknowledges support from Harvard-MIT CUA, NSF Grant No. DMR-1308435, AFOSR Quantum Simulation MURI, ARO MURI Qusim program, AFOSR MURI Photonic Quantum Matter. M.K. acknowledges support from the Technical University of Munich - Institute for Advanced Study, funded by the German Excellence Initiative and the European Union FP7 under grant agreement 291763, and from the DFG grant No. KN 1254/1-1.
\end{acknowledgments}

%

\end{document}